\documentclass[letterpaper,twocolumn,10pt]{article}
\usepackage{usenix2023_SOUPS}
\usepackage{booktabs}
\usepackage{nopageno}

\usepackage{tikz}
\usepackage{amsmath}
\begin{document}

\date{}

\title{\Large \bf Understanding Professional Needs to Create Privacy-Preserving and Secure Emergent Digital Artworks}

\def\plainauthor{Kathryn Lichlyter, Urvashi Kishnani, Kate Hollenbach, Sanchari Das}

\author{
{\rm Kathryn Lichlyter}\\
University of Denver
\and
{\rm Urvashi Kishnani}\\
University of Denver
\and
{\rm Kate Hollenbach}\\
University of Denver
\and
{\rm Sanchari Das}\\
University of Denver
}

\maketitle
\thecopyright

\begin{abstract}

In recent years, immersive art installations featuring interactive artworks have been on the rise. These installations are an integral part of museums and art centers like selfie museums, teamLab Borderless, ARTECHOUSE, and Meow Wolf. Moreover, immersive art have also been increasingly incorporated into traditional museums as well. However, immersive art requires active user participation and often captures information from viewers and participants through cameras, sensors, microphones, embodied interaction devices, surveillance, and kinetic mirrors. Therefore, we propose a new line of research to examine the security and privacy postures of immersive artworks. In our pilot study, we conducted a semi-structured interview with five experienced practitioners from either the art ($2$) or cybersecurity ($3$) fields. Our aim was to understand their current security and privacy practices, along with their needs when it comes to immersive art. From their responses, we created a list of security and privacy parameters, such as, providing opt-in mechanics for data collection, knowledge of data collection tools such as proximity sensors, and creating security awareness amongst participants by communicating security protocols and threat models. These parameters allow us to build privacy-preserving, secure, and accessible software for individuals working in media arts, who often have no background on security and privacy. In the future, we plan to utilize these parameters to develop software in response to those needs and then host an art exhibition of immersive artworks utilizing the platform.


\end{abstract}

\section{Introduction}


Immersive art installations represent a unique genre of artistic expression, characterized by audience engagement and interactivity. In the year $2019$, the immersive art industry was valued at~\textdollar61.8 billion and has only grown since~\cite{basastudioLookAlso}. Within these installations, audiences can become integral components, contributing to the artwork's dynamic narrative through their participation. Some installations even incorporate data collected from participants, seamlessly integrating it into the artistic experience. For instance, notable examples include installations that scan, interpret, and provide feedback on various types of participant data, spanning from physiological responses to environmental interactions. Design I/O~\cite{designioDesignInteractive} interactive installations utilize depth sensors that can generate 3D models of surrounding people and objects~\cite{weiss2015generating}.

The utilization of technology to capture and utilize audience data underscores the importance of addressing security concerns surrounding data collection and storage. As immersive art installations continue to evolve in complexity and scope, ensuring the security and privacy of collected data becomes imperative. Interactive artworks utilize microphones, cameras, sensors, and other devices that can collect data on individuals in real-time~\cite{grenader2015videomob, deliyannis2016interactive}. Several artworks, such as those by Golan Levin~\cite{levin2000painterly, levin2004situ}, Kyle McDonald~\cite{mcdonald2010only}, and Lauren McCarthy~\cite{mccarthy2015getting}, operate by utilizing audiovisual recordings of participants. Other artworks, such as those by Rafael Lozano-Hemmer~\cite{lozano2010rafael}, Teague McDaniel~\cite{teaguemcdaniel}, and Sam Lavigne and Tega Brain~\cite{wu2020sam}, utilize personal data collected from participants. Such data collection practices may be unknown to audiences and considered a breach of privacy by some.

This study applied our prior proposal~\cite{das2022proposal} and aimed to delve deeper into the current approaches, practices, and concerns regarding the cybersecurity of immersive art installations. By interviewing experts from both the cybersecurity and mixed media artistry industries, we seek to gain insights into their methodologies for securing data within their respective domains. Thus, this research endeavors to explore the following research questions:

\begin{itemize}
\item \textbf{RQ1:} In what ways do emergent digital art practitioners consider and prioritize security and privacy in creating immersive artworks?
\item \textbf{RQ2:} What technological needs do installation artists have to create privacy-preserving and secure artworks?
\item \textbf{RQ3:} Through the perspectives of prior research, expert recommendations, and audience viewpoints, what are the primary privacy and cybersecurity concerns associated with immersive digital arts?
\end{itemize}

By synthesizing insights from cybersecurity and media arts experts, as well as considering audience perspectives, this research aims to elucidate the major themes, concerns, and recommendations regarding security and privacy of individuals immersed in interactive art installations.


\section{Method}
To answer our research questions, we first recruited experienced practitioners from both the arts and the cybersecurity field using a two-mode approach. Then, we interviewed the participants to further understand their security and privacy perceptions regarding immersive arts. From the interview transcripts, we identified various themes as part of our analysis. To be eligible to participate in the study, participants had to be either media artists or cybersecurity experts and meet the following requirements: participants should have five or more years of experience in their respective field; participants should be U.S. residents; additionally for media artists, participants should have a portfolio that includes interactive artwork installations, especially installations that collected and/or displayed audience data.

Our recruitment took place via two modes: snowball sampling and a publicly distributed screening survey. We used snowball sampling by emailing local artists and cybersecurity experts with the required experience and asking if they would be interested in participating in our study. We obtained their contact information through their online portfolios. We asked those who were interested to refer us to others in their field, who they knew would be eligible to participate for this study. Additionally, we released a screening survey on social media, inviting interested individuals with five or more years of experience in either art or cybersecurity to fill it out. We contacted respondents who successfully completed the survey and met our study requirements to schedule interviews.

All participants who agreed to participate as a result of snowball sampling and those who were contacted from the survey were asked to schedule time with the research assistant for an interview. To better understand the participants' experiences in their respective fields before the interview, we asked the participants for either a LinkedIn profile or online portfolio that we could review in advance. The interviews were conducted remotely over Zoom~\cite{zoom}. During the interview, we asked questions regarding the participants' experiences in their respective fields, their personal experiences with interactive art installations (either building them or interacting with them), and what security and privacy concerns they had with the interactive art installations they have built and/or experienced. Each interview was video recorded, after obtaining participants' permission through verbal consent. Each interview lasted about an hour. After the interview, each participant was awarded a~\textdollar100 gift card for their time. The transcript for each video was generated using the Temi transcription tool~\cite{temi}. All videos and transcripts were stored on the University of Denver OneDrive server.

To analyze the interviews, the transcripts were open-coded by the researchers to recognize overall common themes among artists and cybersecurity experts. These themes were divided into three main categories: (i) pain points or concerns about the data collection by interactive art installations; (ii) trends in how people educated themselves or taught others about cybersecurity practices; and (iii) trends in software used to build secure technology.

\section{Results}
A total of \textit{five} participants, comprising two media artist and three cybersecurity experts, were interviewed. The participants' backgrounds varied, yet all had encountered interactive art installations in some capacity during their professional careers. Notably, one participant had direct involvement in building such installations, while another had contributed to research on interactive technologies such as virtual reality (VR). The remaining participants had professional ties to the field through collaboration with colleagues or mentoring individuals involved in interactive media.

\subsection{Immersive Art Data Collection Concerns}

Regarding data collection within interactive installations, participants identified several common types of data, including video, audio, voluntarily supplied information such as survey responses, location data, and phone numbers. In the context of interactive artworks, this data collection is essential for understanding audience engagement and enhancing the interactivity of the installations. However, while some data were voluntarily provided by audience members, others, such as data provided by ``proximity sensors," were collected without explicit consent. Participants underscored the importance of implementing opt-in systems to ensure transparency and user consent. They highlighted the complexity of distinguishing between opt-in and opt-out scenarios, acknowledging that current opt-in systems are often intentionally difficult to understand and navigate. This complexity is particularly problematic in interactive artworks, where the seamless and immersive experience should not be compromised by convoluted consent processes. Moreover, participants emphasized the need to collect only necessary data and ensure compliance with relevant data protection regulations such as the General Data Protection Regulation (GDPR) and the California Consumer Privacy Act (CCPA).

Despite the inherent challenges in achieving complete security within interactive installations, participants advocated for a balanced approach that acknowledges the impossibility of eliminating all risks. This approach involves proactive threat modeling, risk acceptance, and a duty of care to protect audience members from potential harm arising from data misuse. They stressed that interactive artworks, which often employ cutting-edge technology and innovative user interactions, present unique cybersecurity challenges. One of the participant emphasized:
\begin{quote}
``There should be a `duty of care' involved to protect [the audience] as much as possible... taking into account all the different ways that [the data] could be misused to harm them."
\end{quote}

By integrating better cybersecurity measures tailored to the specific needs of interactive artworks, creators can ensure that their installations not only provide engaging and immersive experiences but also protect the privacy and security of their audiences. Participants advocated for continuous education and adaptation to evolving security requirements and best practices, recognizing the dynamic nature of both technology and the threats it faces.

\subsection{Continuous Education for Multiple Interdisciplinary Stakeholders}
Participants also emphasized the importance of engaging space managers, such as gallery employees, and implementing guidelines to secure audience data and minimize employee touch-points with interactive installations. They suggested training programs for gallery staff to ensure they understand the importance of data security and are equipped to handle any issues that may arise. These training programs could cover topics such as recognizing potential security threats, responding to data breaches, and maintaining the integrity of the installations. In addition, participants recommended the development of clear policies and procedures for data handling and security, tailored to the specific needs of interactive art installations. In terms of staying abreast of evolving security requirements and practices, participants highlighted the importance of adaptability and continuous learning. Collaborative spaces, including workshops, lectures, and online communities, were cited as valuable resources for exchanging knowledge and staying informed about emerging security trends and solutions. These collaborative efforts can foster a culture of continuous improvement and innovation in cybersecurity practices within the interactive art community. Participants noted that these collaborative spaces provide opportunities for cross-disciplinary learning, where artists can gain insights from cybersecurity experts and vice versa.

Moreover, most cybersecurity experts expressed a keen interest in deepening their understanding of data security issues and solutions specific to interactive art installations. Many admitted to previously overlooking or underestimating the data security implications of the installations they encountered, indicating a growing awareness and interest in this domain. This increased awareness was often spurred by recent high-profile data breaches and privacy scandals, which underscored the importance of integrating security measures into all technological implementations, including art installations. Participants recognized that the evolving nature of cybersecurity threats requires a dynamic and responsive approach, with continuous education and adaptation being key components.

\subsection{Secure and Privacy-focused Software for Immersive Art}

A recurring theme among participants was the concern for better cybersecurity measures tailored to the specific needs of interactive artworks. Participants emphasized the need for software that is accessible and aligns with the project's contextual requirements while ensuring an appropriate level of security and privacy. One participant highlighted their preference for using \textit{Signal}, an end-to-end encrypted communication platform, citing its ease of use and adaptability to various technological contexts. In the context of interactive artworks, they stressed the importance of secure communication among collaborators to protect sensitive creative discussions and proprietary project details for which messaging apps such as Signal can be used. They emphasized the significance of considering factors such as file size and internet bandwidth when selecting communication platforms to ensure accessibility for all collaborators when it comes to artwork discussions. The participant articulated their approach, stating:
\begin{quote}
``How difficult is it for them in the region that they're in to access the [software] that I'm recommending? … If it's a really large file to download, will it download across 3G or 4G? Will it work on the kind of phone [other collaborators] have?"
\end{quote}

\section{Discussion}
The findings of this study illuminate the nuanced approaches adopted by both cybersecurity experts and media artists in addressing data security concerns within the realm of interactive art installations. Through a comprehensive analysis of the data, several key themes and insights emerge, shedding light on the intricate interplay between technology, creativity, and security. This study underscores the dynamic nature of cybersecurity and artistic practices, characterized by continual adaptation and innovation. Both cohorts exhibit a proactive stance towards technological security, emphasizing the importance of minimizing data collection, ensuring secure storage, and effectively communicating security protocols with participating audiences. This multifaceted approach reflects a shared commitment to protecting privacy while facilitating engaging interactive experiences.

One prominent theme is the emphasis on the importance of legal frameworks and compliance standards. The influence of regulations such as the GDPR or CCPA is evident in the practices of both cybersecurity experts and media artists. These legal frameworks provide a strategic imperative for assessing and mitigating risks associated with data storage and handling~\cite{adhikari2023evolution}. While cybersecurity experts caution against the illusion of absolute security, they advocate for a pragmatic approach to risk management and risk acceptance. This approach acknowledges the impossibility of eliminating all risks while striving to minimize potential vulnerabilities. A consensus among study participants revolves around the adoption of protective data handling strategies, particularly advocating for a more transparent and user-centric approach. The preference for an opt-in model for data interpretation and storage underscores a commitment to user privacy and autonomy~\cite{walsh2021my, markert2023transcontinental}. By prioritizing minimalistic data collection and empowering interaction participants to make informed choices regarding their data, artists may seek to foster a culture of responsible data handling within interactive art contexts. This practice not only promotes trust and transparency in the digital realm but also aligns with ethical considerations in data security and honors privacy regulations that impact the users.

Another significant insight is the role of experiential learning in enhancing understanding and awareness of data security practices. Participants express a preference for hands-on experiences, citing interactive art installations as potent educational tools for demystifying concepts of data storage and security. These installations serve as immersive environments where audiences can engage with security concepts in a tangible and intuitive manner. Such initiatives not only enrich the artistic experience but also contribute to broader conversations surrounding cybersecurity best practices. This underscores the transformative potential of art as a vehicle for knowledge dissemination and social change. The study also highlights the importance of continuous education and adaptation to evolving security requirements. Both cybersecurity experts and media artists recognize the dynamic nature of technology and the associated security threats. Collaborative spaces, including workshops, lectures, and online communities, are cited as valuable resources for exchanging knowledge and staying informed about emerging security trends and solutions. These collaborative efforts foster a culture of continuous improvement and innovation in cybersecurity practices within the interactive art community. Participants noted that these collaborative spaces provide opportunities for cross-disciplinary learning, where artists can gain insights from cybersecurity experts and vice versa.

Moreover, the study reveals a growing awareness among cybersecurity experts of the specific data security issues and solutions related to interactive art installations. Many experts admitted to previously overlooking or underestimating the data security implications of the installations they encountered. This increased awareness is often spurred by recent high-profile data breaches and privacy scandals, which underscore the importance of integrating security measures into all technological implementations, including art installations. The evolving nature of cybersecurity threats requires a dynamic and responsive approach, with continuous education and adaptation being key components.

\section{Conclusion}
The synergy between interactive artworks and cybersecurity is essential for creating safe and engaging experiences for audiences. By implementing enhanced security measures, transparent consent systems, and ongoing education and collaboration, the interactive art community can effectively address the unique challenges posed by data collection and privacy concerns. Our work with five professionals in cybersecurity and art highlights the necessity for a proactive and balanced approach to ensure that the creative and experiential aspects of interactive art are preserved while enhancing the privacy and security of audience members. By examining the perspectives and practices of cybersecurity experts and media artists, this study offers a nuanced understanding of the challenges and opportunities inherent in reconciling artistic expression with data security imperatives. This understanding paves the way for informed discourse and collaborative action in protecting privacy and promoting responsible innovation in the ever-evolving digital age. The insights gained from this study emphasize the importance of continuous adaptation and learning, fostering a culture of responsible data handling and ethical practices within the interactive art community. Through these efforts, the interactive art space can continue to thrive as a dynamic and secure environment, benefiting both creators and audiences alike.

\section{Acknowledgement}
This work was supported by the CAHSS Dean’s Award for Interdisciplinary Studies (DAIS) and the Faculty Research Fund for Research for the Privacy and Security of Digital Arts. We would also like to thank our participants for this study and acknowledge the Inclusive Security and Privacy-focused Innovative Research in Information Technology (InSPIRIT) Lab at the University of Denver for supporting this work. Any opinions, findings, conclusions, or recommendations expressed in this material are solely those of the authors.

\bibliographystyle{plain}
\bibliography{ref}

\end{document}